\documentclass[mathleft
]{an}
\usepackage{graphicx}
\usepackage{times}
\begin{document}

\Pagespan{789}{}
\Yearpublication{2012}%
\Yearsubmission{2012}%
\Month{11}%
\Volume{999}%
\Issue{88}%

\title{New results in RR\,Lyrae modeling: convective cycles, additional modes and more}

\author{L. Moln\'ar\inst{1}\fnmsep\thanks{\email{molnar.laszlo@csfk.mta.hu}},
Z. Koll\'ath\inst{1}, R. Szab\'o\inst{1}, E. Plachy\inst{2}
}
\titlerunning{New results in RR Lyrae modeling}
\authorrunning{L. Moln\'ar et al.}
\institute{
Konkoly Observatory, MTA CSFK, Konkoly Thege Mikl\'os \'ut 15-17, H-1121, Budapest, Hungary
\and 
Department of Astronomy, E\"otv\"os University, P\'azm\'any P\'eter s\'et\'any 1/A, H-1117, Budapest, Hungary}
\received{}
\accepted{}
\publonline{later}

\keywords{RR Lyrae stars -- hydrodynamics -- convection}

\abstract{Recent theoretical and observational findings breathed new life into the field of RR\,Lyrae stars. The ever more precise and complete measurements of the space asteroseismology missions revealed new details, such as the period doubling  and the presence of the additional modes in the stars. Theoretical work  also flourished: period doubling was explained and an additional mode has been detected in hydrodynamic models as well. Although the most intriguing mystery, the Blazhko-effect has remained unsolved, new findings indicate that the convective cycle model can be effectively ruled out for short- and medium-period modulations. On the other hand, the plausibility of the radial resonance model is increasing, as more and more resonances are detected both in models and stars. }

\maketitle

\section{Introduction}
RR\,Lyrae stars are one of the oldest class of pulsating variables, with a rich history that spans more than a century (Kolenberg 2012). Many details are well understood about them: they are old, helium-burning stars, pulsating in simple radial modes, either in the fundamental mode or the first overtone or in both of them. A few unsolved or not fully understood issues remain though, most notably the Oosterhoff-dichotomy (Oosterhoff 1939) and the Blazhko-effect. The latter, the (quasi-)periodic modulation of the pulsation amplitude and phase is one of the most stubborn problems of stellar astrophysics, first observed a century ago by Blazhko (1907) and Shapley (1916). 

Radial pulsation allows for one-dimensional, non-linear, non-adiabatic hydrodynamic modeling of RR\,Lyrae stars. Current, state-of-the-art models---such as the Florida-Budapest code, used by our group---incorporate the effects of turbulent convection as well (Koll\'ath \& Buchler 2001; Koll\'ath et al. 2002). The code first creates an equilibrium model and calculates the linear eigenfrequencies and growth rates of the model. This is then perturbed and followed non-linearly as the pulsation approaches the full amplitude. The model can be iterated effectively into the limit cycle solution with the relaxation method (Stellingwerf 1974), and the stability of the limit cycle can be addressed (Floquet 1883). In this paper, we summarize some of the recent results our group obtained with the Florida-Budapest code, and the synergies with the newest observational data, especially with the \textit{Kepler} mission (Borucki 2010).

\section{Convective cycles: the Stothers-model}
Neither of the two classical proposals for the Blazhko-effect, the magnetic oblique rotator (Cousens 1983; Shibahashi 2000) and the nonradial resonant rotator models (Dziembowski \& Cassisi 1999; Nowakowski \& Dziembowski
2001) can be modeled with a one-dimensional, non-magnetic code. However, another idea has (re)surfaced recently by Stothers (2006): the model of convective cycles proposes that periodic buildup and breakdown of turbulent magnetic fields in the convective envelope can alter the stellar structure sufficiently to create the observed variations in the pulsation. 

Although a detailed modeling effort would require a pulsating, 3D, MHD model, the  main idea, \textit{i.e.} the variations of the convective properties can be included into the existing 1D codes reasonably well. In this case the cause of the modulation is an \textit{ad hoc}, manually inserted variation in one or more of the internal convective parameters, such as the famous mixing length parameter. This way, Smolec et al. (2011) showed that very strong internal variation was required to reproduce the modulation of the star RR\,Lyr. 

Our calculations focused on the dynamics of the process (Moln\'ar, Koll\'ath \& Szab\'o 2012a). The results confirmed our initial concerns: planting the same internal modulation into a stellar model created very different modulation in the observable quantitites (luminosity, radius), depending on the modulation periods.  We concluded that the Stothers-model is not effective under a Blazhko-period shorter than about 100-150 days. This runs counter to the fast and strong modulation observed in several stars: \textit{e.g.} 16.5 days for MW\,Lyr (Jurcsik et al. 2008), 53.1 days for V445\,Lyr (Guggenberger et al. 2012) or 39.6 days for RR\,Lyr itself (Kolenberg et al. 2011).

\section{Additional modes}
Recent observations, mainly from CoRoT and \textit{Kepler}, revealed that RR\,Lyrae stars are not limited to the two lowest radial modes and the three classical (RRab, RRc and RRd) states (Benk\H{o} et al. 2010). Additional modes, small, mmag-range signals in the Fourier-spectra, outside the usual patterns, were observed in all classes: RRab (Poretti et al. 2010; Benk\H{o} et al. 2010; Guggenberger et al. 2011, 2012), RRc (Moskalik et al. 2013) and RRd stars (Gruberbauer et al. 2007; Chadid 2012). They were detected in ground-based, multicolor photometry as well (S\'odor et al. 2012). However, the explanation of the additional signals, even if they coincide with the expected frequency of the first or second overtone, has been challenging. Another important aspect is that apart from a single star (V350 Lyr, Benk\H{o} et al. 2010), all the other RRab stars with additional mode(s) show at least hints of modulation as well, strongly suggesting a connection between the two phenomena.

\subsection{The first surprise: period doubling}
A few interesting model runs turned up during the very first model runs connected to the Stothers-model. Some showed alternating higher and lower maxima, and a few others varied irregularly. The former behavior is known as period doubling but it had not been observed in RR\,Lyrae stars previously. Shortly after, however, the \textit{Kepler} telescope started its operation, and it turned out that some stars, RR\,Lyr among them, indeed show period doubling (Kolenberg et al. 2010; Szab\'o et al. 2010). These data confirmed that a legitimate new dynamical behavior was observed in the models and triggered extensive numerical calculations (Koll\'ath, Moln\'ar \& Szab\'o 2011). The models revealed that a previously unexpected, high order (9:2) resonance between the fundamental mode and the ninth radial overtone is responsible for the phenomenon. This was the first time that the presence of an additional mode was confirmed in RR\,Lyrae stars.

The 9:2 resonance was also modeled with amplitude equations (Buchler 1993) by Buchler \& Koll\'ath (2011). The calculations revealed that the resonance is capable to create not only period doubling but also periodic and chaotic modulation. Though hydrodynamic RR\,Lyrae models have not been able to reproduce this modulation yet, the radial resonance model gained popularity, thanks to recent developments: the criticism of the Stothers-model (Section 2), the detection of additional modes and the broad range of possible resonances and mode interactions (Sections 3.2-3.4), and the detection of modulation in period-doubled BL\,Herculis models (Smolec \& Moskalik 2012).

\begin{figure}
\includegraphics[scale=0.58]{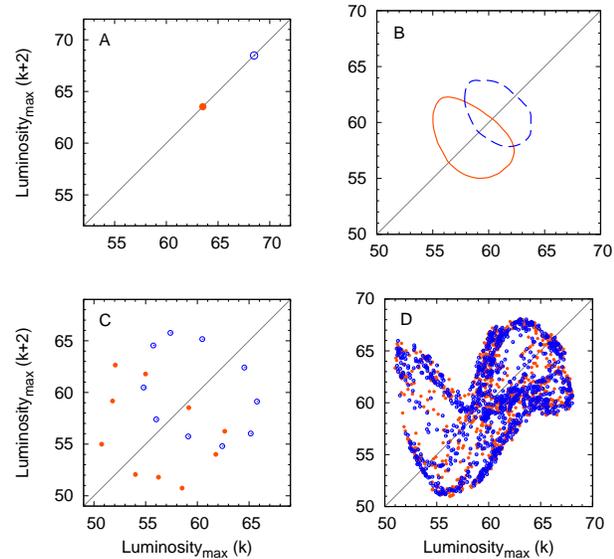}
\caption{Return maps of four different dynamical states of RR\,Lyrae models. Here we use the($k$+2)th  vs. $k$th maxima instead of the successive ones to separate the higher and lower cycles (orange dots/continuous line and blue circles/dashed line) of period doubling. \textit{A}: period doubling only, the maxima alternate between the two values. \textit{B}: non-resonant three-mode state. If two different modes are present (the fundamental and the first overtone, the 9th overtone is hidden in these cases), the system travels around a loop, according to the relative phase of the two modes. The fundamental mode is also period-doubled, so the system alternates between the two loops. \textit{C}: a resonant three-mode state. Here $P_0$ and $P_1$ are commensurate and the two variations return to the same phase values after a few (in this case, twenty) pulsation cycles. \textit{D}: a chaotic state. In this case, the alternation of even and odd cycles is not strict any more but experiences reversals, hence the mixing of blue and orange symbols. }
\label{retmaps}
\end{figure}

\subsection{The second surprise: additional modes and chaos}
Period doubling can lead to chaos through a bifurcation cascade in dynamical systems. Two model solutions with irregularly varying maxima and minima were indeed found to be chaotic (Plachy, Koll\'ath \& Moln\'ar 2013). However, we could not identify any period-doubling cascade that led to those solutions. Closer inspection of the models revealed that in fact they contain not only the fundamental mode and the ninth overtone, but the first overtone as well. This bears a strong similarity with some of the \textit{Kepler} RR\,Lyrae stars where both period doubling and a first overtone peak was observed. The first results indicate that the presence of period doubling plays a crucial role in the stability of the first overtone, and the three-mode state is a natural extension of the two-mode, period-doubled one. However, as the number of modes involved grows, the variety of dynamical states also expands. If two modes are present in RR\,Lyrae models, the result is either the classical double-mode (RRd) pulsation or the period doubling (and higher-order bifurcations, see Koll\'ath et al. 2011). If three modes are present, we observe a great variety of non-resonant, resonant (incommensurate or consummerate $P_0$ and $P_1$ values) and chaotic states, along with bifurcation cascades. These new findings will be presented in a separate paper (Koll\'ath et al. in prep.).

\begin{figure}
\includegraphics[scale=0.70]{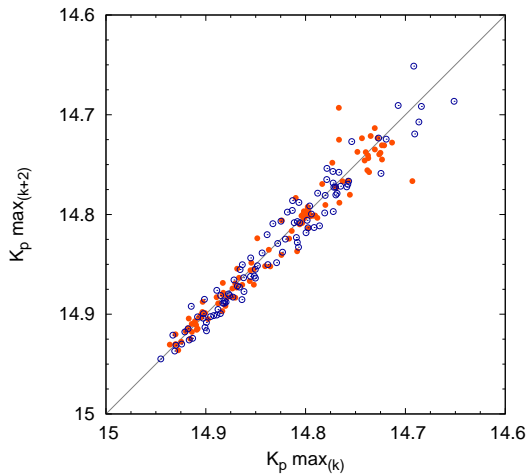}
\caption{Return map from the first two quarters of \textit{Kepler} observations of the star V808\,Cyg. The dominant feature is the amplitude modulation from the Blazhko-effect that smears out any pattern along the diagonal. Blue circles and orange dots are the same as in Figure \ref{v808}.}
\label{v808_rm}
\end{figure}

\subsection{Return maps and maxima}
Differentiating between the states of the three-mode solutions was initially a challenge: the time series of both the non-resonant and chaotic models look irregular, while the Fourier-spectrum is always complicated because of the linear combinations, especially with the strong harmonics of the fundamental frequency. We used two different techniques to determine the dynamical state of the models. One is the return map (also called first return map): we select a well-defined state of the light curve, \textit{e.g.} the maximum brightness at every pulsation cycle, and plot every value against the preceding one. Given enough cycles to populate the map, its structure then may tell us the dynamical state of the system at a glance. We present four examples on Figure \ref{retmaps}: subplot \textit{A} shows a period-doubled, \textit{B} and \textit{C} a three-mode (doubled fundamental plus the first overtone) model with non-resonant and resonant period ratios, while \textit{D} shows a very characteristic, scatter-plot-like map of a chaotic system. 

\begin{figure*}
\includegraphics[scale=0.80]{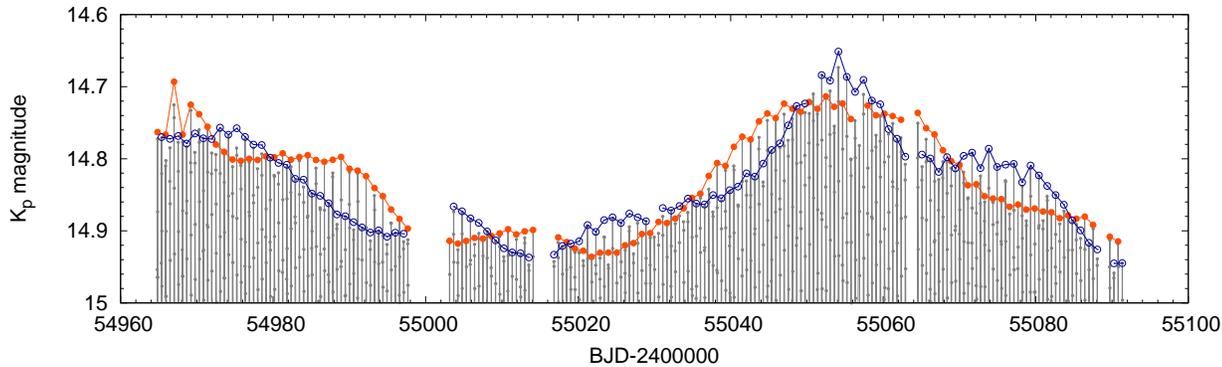}
\caption{Brighter part of the light curve (grey) and maxima of the pulsation cycles of V808\,Cyg from the first two quarters of \textit{Kepler} long cadence observations (Szab\'o et al. 2010). Larger blue circles and orange dots represent maximum brightness of even and odd cycles, respectively, in the same way as in RR\,Lyr in Moln\'ar et al. (2012b). The higher and lower cycles of period doubling clearly reverse, on shorter timescales as the modulation period. The additional scatter in the maxima is likely an undersampling effect: long cadence (29.4--min sampling) data is not well suited for capturing the sharp maxima of the pulsation, so the determination of maximum brightnesses, especially at high Blazhko-phase, can be challenging.}
\label{v808}
\end{figure*}

Return maps are not always the best solution to visualize the system, however. They work well with the models where the mode amplitudes are usually constant but almost all additional-mode stars show the Blazhko-effect too. The amplitude modulation smears the patterns up and down along the diagonal, as in the example case of V808\,Cyg (Figure \ref{v808_rm}). The other method we use is to inspect the maxima themselves, separating  the odd and even cycles, hence separating the lower- and higher-amplitude branches of period doubling. The result is shown in Figure \ref{v808}: we used the same data and maxima as in  Szab\'o et al. (2010), but contrary to the implications in that paper, period doubling never really switches off, it only reverses or swaps branches, when the amplitude of PD decreases. The timescale of the reversal is faster than the Blazhko-cycle (8-18 days vs. $\approx 90$ days). Very similar behavior was observed in RR\,Lyr itself too (Moln\'ar et al. 2012b).

\subsection{Resonances and chaos}
The 9:2 resonance that creates the period doubling was already regarded as quite unusual for pulsating stars. The three-mode models allow for even higher ratios: considering the usual $P_0/P_1 \sim 0.72-0.75$ period ratios for the fundamental mode and the first overtone, the following resonances may exist: 3:4, (6:8), 8:11, 14:19, 20:27, etc. We observed several of these ratios in the three-mode models, up to very high integers but they will also be discussed in a separate paper, dedicated to the three-mode models (Koll\'ath et al., in prep.). Very interestingly, we have not been able yet to find the lowest, 3:4 resonance case (which should be called 6:8 as period doubling requires even number of fundamental-mode cycles to repeat). We did find a temporary state---and the first overtone itself---in the \textit{Kepler} observations of RR\,Lyr, however, where the star does in fact seem to approach the vicinity of the 6:8 resonance (Moln\'ar et al. 2012b). 

A possible extension to the hydrodynamics is the global flow reconstruction method (GFRM, Serre, Koll\'ath \& Buchler 1996) applied to the chaotic solutions. The GFRM fits polynomials to the data and creates a series of maps. Then these maps are iterated from different initial values. The maps and the synthetic signals (iterated time series) are very close to each other in the phase space, similarly to the chaotic and non-chaotic states of the models. However, the GFRM maps can cover a larger portion of the same phase space, compared to the hydrodynamic calculations. We were able to find synthetic signals where the two main periodicites are in or very close to the 3:4 resonance. The results will also be presented in a dedicated paper (Plachy et al., in prep.).

\section{Future work}
Though our understanding of RR\,Lyrae stars has developed greatly, there are still several areas that need further attention. 

Period doubling in the stars is not a stable phenomenon: the amplitude is changing and the order of higher and lower cycles reverses. This has been observed---along with Blazhko-like modulation---in models of BL\,Herculis-type stars by Smolec \& Moskalik (2012) but not yet in RR\,Lyrae models.

RR\,Lyrae stars with additional modes more often show the signs of the second or both the first and second overtone than the first overtone only (Benk\H{o} et al. 2010). In contrast, hydrodynamic models have only explained the presence of the latter mode. A three-mode resonance of $3 f_0 + f_2 = f_9$, based on linear model results, was proposed (Moln\'ar, Koll\'ath \& Szab\'o 2013), but it was not detected in nonlinear models yet.

First-overtone stars are also worth attention: all four RRc stars known in the \textit{Kepler} field show additional modes (Moskalik et al. 2013). A detailed, dedicated study of these stars is under way (Moskalik et al., in prep.). The multi-mode regimes also have to be incorporated into the classical modal selection diagram (Szab\'o, Koll\'ath \& Buchler 2004) along with the period-doubled cases. 

\textit{Kepler} revealed that the properties of the Blazhko-modulation can be extremely diverse among otherwise similar stars (Benk\H{o} et al. 2010). Investigating the modulation characteristics of the sample both individually and comparatively might provide us more clues about the mechanism(s) behind the Blazhko-effect.

\acknowledgements
This work was supported by the Hungarian OTKA grants K83790 and K81421, the HUMAN MB08C 81013 grant of the MAG Zrt.\, and the `Lend\"ulet-2009' Young Researchers' Programme. RSz acknowledges the Bolyai J\'anos Scholarship. The European Union and the European Social Fund have provided financial support to the project under the grant agreement no. T\'AMOP-4.2.1/B- 09/1/KMR-2010-0003.

\end{document}